# A Novel Elliptic curve cryptography Processor using NoC design

Hamid Javashi[1] and Reza Sabbaghi-Nadooshan[2]

[1] Electronic Engineering Department, Islamic Azad University Central Tehran Branch,
Tehran, Iran
*hamid.jav@gmail.com*

[2] Electronic Engineering Department, Islamic Azad University Central Tehran Branch,
Tehran, Iran
*R_sabbaghi@iauctb.ac.ir*

**Abstract**
In this paper, we propose an elliptic curve key generation processor over $GF(2^m)$ and $GF(P)$ with Network-on-Chip (NoC) design scheme based on binary scalar multiplication algorithm. Over the Two last decades, Elliptic Curve Cryptography (ECC) has gained increasing acceptance in the industry and the academic community. This interest is mainly caused by the same level of security with relatively small keys provided by ECC comparing to large key size in Rivest Shamir Adleman (RSA). Parallelism can be utilized in different hierarchy levels as shown in many publications. By using NoC, a new method with the reduced latency of point multiplication (with parallel field arithmetic) is introduced in this paper.

*Keywords: ECC, ECPM, NoC, parallelism.*

## 1. Introduction

Elliptic Curve Cryptography (ECC) has been proposed by Miller [1] and Koblitz [2] in the mid 1980s.
A flexible architecture for Elliptic Curve Point Multiplication (ECPM) with the help of Network on Chip (NoC) over $GF(p)$ and $GF(2^m)$ is presented in this paper. ECPM is the basic operation of any elliptic curve cryptosystem and it is essential that it is implemented efficiently. Several implementations of ECPM have been introduced e.g. in [3-5] and one implementation with NoC design is proposed in [6].
Elliptic curve cryptography (ECC) is a public key cryptography system which attractive primarily because of its shorter wordlengths compared to RSA implementations with the same level of security [7]. 2048-bit RSA, for example, can be replaced by an ECC system of equivalent security while reducing the key wordlength by a factor of eight [8]. Some benefits of having smaller key sizes include reductions in processing power, storage space, and bandwidth Due to these many advantages of ECC, a number of software [9, 10] and hardware [11–14] implementations have been proposed. ECC is widely deployed in many applications such as wireless device [15], digital signal processing [16], RFID tag [17], and smart cart [18].

For implementations of ECC, finite fields $GF(p)$ and $GF(2^m)$ have been used, where p is a prime and m is a positive integer greater than one. In particular $GF(2^m)$, which is an m-dimensional extension field of $GF(p)$, is suitable for hardware implementations because there is no carry propagation in arithmetic operations.

A relatively large amount of research has been performed on hardware implementations of ECCs over $GF(2^m)$, including those given in [19-20]. This is because $GF(2^m)$ implementations are generally faster and more compact than their $GF(p)$ equivalents. However, conventional hardware implementations of ECCs over $GF(p)$ are not very flexible, as the field parameters used are often fixed. In practice, ECC has many different operational modes and primitives and therefore architectures need to be flexible. Thus, $GF(2^m)$ may not be the ideal choice if flexibility is an important requirement. Moreover, RSA, the most widely used public-key cryptosystem uses arithmetic over $GF(P)$.

Therefore, it is attractive to create hardware architectures suitable for ECCs over $GF(P)$, which have the capability of supporting RSA as well as ECC cryptography.
The sequential nature of the point multiplication makes efficient use of parallelization challenging, however although the point multiplication itself is hard to parallelize, it is possible to efficiently use parallelism in point operations.

The rest of the paper is organized as follows: the theoretical background, including ECC arithmetic is described in section 2; section 3 presents the design of ECC general processor, NoC properties which are needed in ECC is presented in section 4. In section 5 we introduce new design of ECC processor on NoC, and finally, Conclusions are drawn in section 6.

## 2. ECC arthimetic

The Weierstraß equation of an elliptic curve E defined over $\overline{K}$ using affine coordinates is:

$$E : y^2 + a_1 xy + a_3 y = x^3 + a_2 x^2 + a_4 x + a_6 \quad Eq(1)$$

where $a_1, a_2, a_3, a_4, a_5, a_6 \in \overline{K}$. An elliptic curve E is then defined as the set of solutions of this equation in the affine plane, together with the point at infinity $O$.

For a non-supersingular curve equation of characteristic 2 (defined over a field $GF(2^m)$), the reduced Weierstraß equation has the form :

$$y^2 + xy = x^3 + ax^2 + b \quad Eq(2)$$

With $a, b \in GF(2^m), b \neq 0$ and the set of point which satisfy Eq. (2).

For a non-singular curve equation defined over a field GF(P) and of characteristic $k \neq 2, 3$ the reduced Weierstraß equation has the form:

$$y^2 \equiv x^3 + ax + b \pmod{p} \quad Eq(3)$$

Where $a, b \in GF(P)$ and

$$4a^3 + 27b^2 \not\equiv 0 \pmod{P},$$

$(x, y) \in GF(P) \times GF(P)$ which satisfy Eq. (3) [21].

### 2-1 Group law

While using elliptic curve, the fundamental operations are the point addition and doubling.
If there are two points on curve namely, P $(x_1, y_1)$, Q $(x_2, y_2)$ and their sum given by point, R$(x_3, y_3)$ the algebraic formulas for point addition and point doubling are given by following equations:

$$\begin{aligned}
&if \quad P \neq Q \\
&x_3 = \left(\frac{y_2 - y_1}{x_2 - x_1}\right)^2 - x_1 - x_2 \\
&y_3 = \left(\frac{y_2 - y_1}{x_2 - x_1}\right)(x_1 - x_3) - y_1 \\
&if \quad P = Q \\
&x_3 = \left(\frac{3x_1^2 + a}{2y_1}\right)^2 - 2x_1 \\
&y_3 = \left(\frac{3x_1^2 + a}{2y_1}\right) - (x_1 - x_3) - y_1
\end{aligned}$$

where the addition, subtraction, multiplication and the inverse are the arithmetic operations over GF(p).

$$\begin{aligned}
&if \quad P \neq Q \\
&x_3 = \left(\frac{y_2 + y_1}{x_2 + x_1}\right)^2 + x_1 + x_2 \\
&y_3 = \left(\frac{y_2 + y_1}{x_2 + x_1}\right)(x_1 + x_3) + y_1 + c \\
&if \quad P = Q \\
&x_3 = \left(\frac{x_1^2 + a}{c}\right)^2 \\
&y_3 = \left(\frac{x_1^2 + a}{c}\right)^2 + (x_1 + x_3) + y_1
\end{aligned}$$

where the addition, subtraction, multiplication and the inverse are the arithmetic operations over GF($2^m$) [21].

### 2-2 Scalar multiplication algorithms

ECPM is defined on E so that :

$$Q = k.P = P + P + \ldots + P \quad Eq(4)$$

where Q and P are distinct points on E and k is a large integer ECPM in Eq. (4) is calculated with successively performed operations called point addition and point doubling. The hierarchy of arithmetics for EC point multiplication is depicted in Fig. 1.

A point addition is an operation $P_3 = P_1 + P_2$, where P, are distinct points on an elliptic curve, and a point doubling is an operation $P_3 = 2P_1$. Several efficient methods to compute ECPM in Eq. (4) have been proposed. The basic method for computing kP is the addition-subtraction method described in draft standard IEEE P1363 [22]. This method is an improved version over the well known double and add (or binary) method, which requires no precomputations. Several proposed generalizations of the binary method (for exponentiation in a multiplicative group), such as the k-ary method, the signed window method, can be extended to compute elliptic scalar multiplications over a finite field [23-24].

A different approach for computing kP was introduced by Montgomery [25]. This approach is based on the binary method.

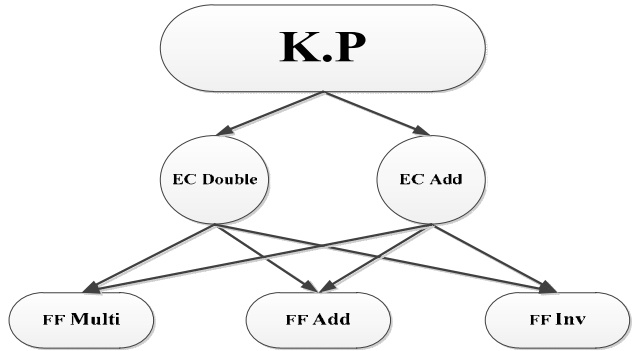

Fig. 1. EC arithmetic hierarchy

In this paper we use binary method for both GF(p) and GF($2^m$), Binary method (or double-and-add method) is probably the most common way to compute Q=k.P .

In the binary method, k is represented with binary expansion as $K = (1, k_{l-1}, \ldots, k_1, k_0)_2$ and a point doubling is performed for each bit $k_i$ and a point addition if $k_i$=1. Thus, because $l \approx m$, $m$ point doublings and $m/2$ point additions are required on average.

```
Binary method :
Input :      P ∈ E, K = (1, k_{l-1}, …, k_1, k_0)_2
Output :     Q = k.P
             Q ← P
             for i from l − 2 down to 0 do
                 Q ← 2Q
                 if k_i = 1
                     then Q ← Q + P
```

This algorithm that mentioned above is same for GF(P) and GF($2^m$), but there is different in finite field arithmetic computation between two fields GF(P) and GF($2^m$).

## 2-3 Projective coordinate

Naturally, point addition and point doubling also require a field inversion when using affine coordinates $(x, y)$. Since inversion is a very expensive operation compared to multiplication, addition and squaring in finite fields, we use projective coordinates. In standard projective coordinates the points on the elliptic curve are represented as a triple $(X, Y, Z)$ in such a way that $x \rightarrow X/Z$ and $y \rightarrow Y/Z$. By using projective coordinates only one finite field inversion is required at the end of a scalar multiplication in order to transform the projective coordinates back to affine coordinates. Several types of projective coordinates have been proposed.

For GF(p) includes, Standard projective coordinates, Jacobian projective coordinates and Chudnovsky coordinates, and for GF($2^m$) includes Standard projective coordinates, Jacobian projective coordinates and finally L´opez-Dahab (LD) projective coordinates [21].

We use Jacobian projective coordinates and and L´opez-Dahab (LD) projective coordinates for arithmetic over GF(P) and GF($2^m$), respectively.

The number of multiplication and addition in scalar multiplication is shown in Table 1.

According to binary method algorithm with projective coordinate, the scalar multiplication is performed over three steps; (i) initialization operations, (ii) iteration of the point addition and point doubling; and (iii) the conversion to affine coordinates.

Table 1: Number of arithmetic computation in binary method

| | L´opez-Dahab (LD) projective coordinates [24] | | | Jacobian projective coordinates [24] | | |
|---|---|---|---|---|---|---|
| | M-Add | M-Double | M-xy | M-Add | M-Double | M-xy |
| ADD | 2 | 1 | 6 | 2 | 1 | 6 |
| MUL | 4 | 2 | 10 | 4 | 2 | 10 |
| INV | 0 | 0 | 1 | 0 | 0 | 1 |
| SQR | 1 | 4 | 1 | 1 | 4 | 1 |

## 3. ECC Processor design

As mentioned above, we will study the scalar multiplication method based on binary method Algorithm. A block diagram of the elliptic curve processor is show in Fig. 2. It consists of an arithmetic logic unit, register bank and controller unit.

Register bank consists of several registers which is stored the information in each stage of binary method. The control module consists of a finite state machine. It generates the control signals for the initialization operations of finite field, the point addition and point doubling operations and the conversion to affine coordinates operations. Finally the arithmetic and logical unit (ALU) allow parallel execution of finite field addition, inversion, squaring and multiplications, which are controlled by the control unit.

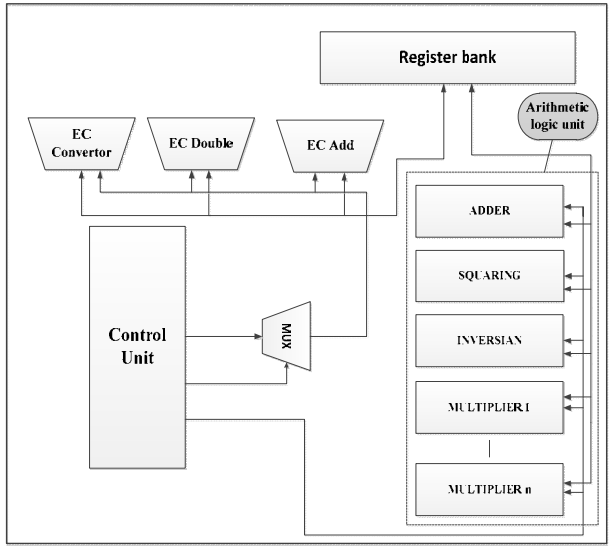

Figure 2: block diagram of the elliptic curve processor

## 4. Network on Chips

The advances in the semiconductor technology and the shrinking feature size in the deep submicron era, have led to an impressive increase in the number of transistors available on a single chip. This huge number of transistors enables us to integrate complex system-on-chip (SoC) designs containing a large number of processing cores together with large amounts of embedded memory and high-bandwidth I/O on a single chip. On the other hand, the applications are becoming more and more complex. The large amount of computational resources together with complicated communication patterns due to complex applications requires higher degrees of support from communication resources. Besides, as technology advances, the delay and power consumption of global interconnect structures (due to a large number of drivers) will be a major bottleneck for SoC design.

Network-on-Chip (NoC) is a promising communication paradigm for multiprocessor system-on-chips. This communication paradigm has been inspired from the packet-based communication networks and aims at overcoming the performance and scalability problems of the shared buses in multi-core SoCs (System on Chips) [26].

Although the concepts of NoC are inspired from the traditional interconnection networks, they have some special properties which are different from the traditional networks. Compared to traditional networks, power consumption is the first-order constraint in NoC design [27]. As a result, not only should the designer optimize the NoC for delay (for traditional networks), but also for power consumption.

The choice of network topology is an important issue in designing a NoC. Different NoC topologies can dramatically affect the network characteristics, such as average inter-IP distance, total wire length, and communication flow distributions. These characteristics in turn, determine the power consumption and average packet latency of NoC architectures.

In general, the topologies proposed for NoCs can be classified into two major classes, namely regular tile-based and application-specific. Compared to regular tile-based topologies, application-specific topologies are customized to give a higher performance for a specific application. Moreover, if the sizes of the IP cores of a NoC vary significantly, regular tile-based topologies may impose a high area overhead. On the other hand, this area overhead can be compensated by some advantages of regular tile-based architectures. Regular NoC architectures provide standard structured interconnects which ensures well-controlled electrical parameters. Moreover, usual physical design problems like crosstalk, timing closure, and wire routing and architectural problems such as routing and switching strategies and network protocols can be designed and optimized for a regular NoC and be reused in several SoCs.

The mesh topology is the simplest and most dominant topology for today's regular tile-based NoCs. It is well known that mesh topology is very simple. It has low cost and consumes low power [28], so we use this topology for ECC.

## 5. ECC with NoCs

Parallelism in point operations is also an efficient way to reduce latency of point multiplication as shown in [29]. Parallelism can be utilized in binary methods. The information exchange rate can be raised when each section in the processor (elliptic curve) is a core on NoC (According to the NoC features). By Placing the core in the right place (Figure 3) in NoC performance is improved and power consumption is reduced. It is noteworthy that the design alongside the ideal processor will have Significant impact on the speed and power consumption.

In this paper for implementation of elliptic curve we needed to 12 cores, so we used 4×3 two-dimensional mesh NoC. Cores such as MUL(multiplying) that use more than others, place on the middle. Cores that use less, place on the corners.

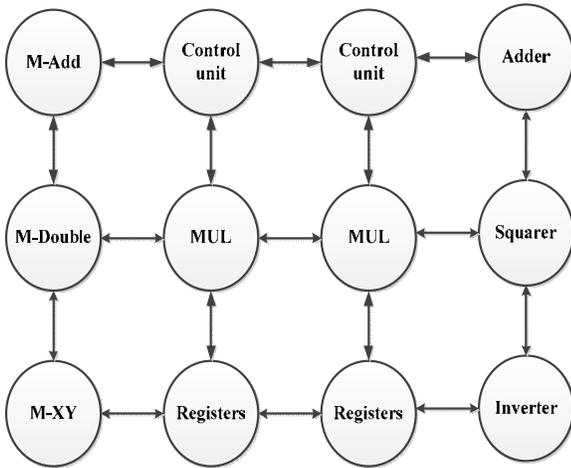

Figure 3. Implementation of ECC by NoC

## 6. Conclusion

In this paper, we introduced configurable NoC architecture for Elliptic Curve Cryptosystem that improved the speed of data exchange between cores (finite field arithmetic unit and control unit) with the help NoC features. Our experience shows that performance in ECC processor designs is dictated mainly by data transfer speed.

**Hamid Javashi** is MS candidate in electronic engineering from Central Tehran Branch of Islamic Azad University.

**Reza Sabbaghi-Nadooshan** received the B.S. and M.S. degree in electrical engineering from the Science and Technology University, Tehran, Iran, in 1991 and 1994 and the Ph.D. degree in electrical engineering from the Science and Research Branch, Islamic Azad University, Tehran, Iran in 2010. From 1998 he became faculty member of Department of Electronics in Central Tehran branch, Islamic Azad University, Tehran, Iran. His research interests include interconnection networks, Networks-on-Chips, Hardware design and embedded systems.